\documentclass[preprint]{aastex}
 
\slugcomment{Accepted for publication in AJ}

\shorttitle{Counterrotating Gas and Stars}
\shortauthors{Kannappan \& Fabricant}
 
\newcommand{\MSun}{$\rm M_{\sun}$}
\newcommand{\LSun}{$\rm L_{\sun}$}

\begin{document}
 
\title{A Broad Search for Counterrotating Gas and Stars:  \\
Evidence for Mergers and Accretion}

\author{S. J. Kannappan and D. G. Fabricant}
\affil{Harvard-Smithsonian Center for Astrophysics,
    Cambridge, MA 02138}
\authoraddr{ Mail Stop 20, 60 Garden St. Cambridge, MA 02138,
        (email: skannappan@cfa.harvard.edu, dfabricant@cfa.harvard.edu)}

\begin{abstract}
We measure the frequency of bulk gas-stellar counterrotation in a
sample of 67 galaxies drawn from the Nearby Field Galaxy Survey, a
broadly representative survey of the local galaxy population down to
M$_{\rm B}\sim$ $-$15.  We detect 4 counterrotators among 17 E/S0's
with extended gas emission ($24^{+8}_{-6}\%$).  In contrast, we find
no clear examples of bulk counterrotation among 38 Sa--Sbc spirals,
although one Sa does show peculiar gas kinematics.  This result
implies that, at 95\% confidence, no more than 8\% of Sa--Sbc spirals
are bulk counterrotators.  Among types Sc and later, we identify only
one possible counterrotator, a Magellanic irregular.  We use these
results together with the physical properties of the counterrotators
to constrain possible origins for this phenomenon.

\end{abstract}
\keywords{galaxies: evolution 
--- galaxies: formation 
--- galaxies: interactions 
--- galaxies: kinematics and dynamics }

\section{Introduction}

Galaxies with counterrotating gas and stars offer dramatic evidence
for the hypothesis of hierarchical galaxy formation.  Counterrotation
highlights the possibility of multiple events in a galaxy's formation
history, as opposed to isolated collapse and infall models
\citep{rubin:multi-spin,schweizer:observational}.  Studies of
elliptical, S0, and Sa galaxies have demonstrated that gas-stellar
counterrotation and related phenomena such as non-coplanar rotation
may be quite common
\citep[e.g.][]{bertola.buson.ea:external,jore:kinematics}, although
cospatial stellar-stellar counterrotation may be less so
\citep{kuijken.fisher.ea:search}.  However, the frequency of these
phenomena in the general galaxy population is uncertain.  Equally
important, we do not know whether the physical processes responsible
for counterrotation merely perturb the host galaxy along the Hubble
sequence, or completely reshape it.  Comparisons of gas accretion and
merger scenarios have been inconclusive
\citep{thakar.ryden.ea:ngc,galletta:counterrotation}.

Combining the compilation of
\citet[review][]{galletta:counterrotation} with a few recent
discoveries
\citep{kuijken.fisher.ea:search,morse.cecil.ea:inclined,jore:kinematics},
we identify in the literature at least 18 elliptical and S0
gas-stellar counterrotators, as well as 9 spiral and later type
galaxies.  Non-coplanar rotation, particularly in the stable
configuration of a polar ring, may also be common:
\citet{whitmore.lucas.ea:new} list 6 kinematically confirmed polar
ring galaxies and many more visual candidates.  Unfortunately the
statistical implications of these numbers are unclear, since so many
of the discoveries were serendipitous.

To date, the best statistical estimates of the frequency of
gas-stellar counterrotation come from studies restricted to early type
galaxies.  \citet{bertola.buson.ea:external} and
\citet{kuijken.fisher.ea:search} have analyzed samples of emission-line S0's
and report that $\sim$20--25\% of these galaxies show counterrotation
or other strong kinematic decoupling between the gas and stars.
\citet{jore:kinematics} has surveyed 23 isolated, unbarred Sa galaxies
and finds 4 that show clear gas-stellar counterrotation along the
major axis, with the bulk of the stars rotating opposite to the bulk
of the gas \citep[see also][]{haynes.jore.ea:kinematic}.

Here we report the frequency of gas-stellar counterrotation along the
major axes of 67 galaxies ranging from ellipticals to Magellanic
irregulars.  This work makes use of data from the recently completed
Nearby Field Galaxy Survey
\citep[NFGS,][]{jansen.franx.ea:surface,jansen.fabricant.ea:spectrophotometry,
kannappan.fabricant.ea:kinematics}, which contains imaging,
spectrophotometric, and kinematic observations for $\sim$200 galaxies
drawn from the CfA~1 redshift survey \citep{huchra.davis.ea:survey}.
\citeauthor{jansen.franx.ea:surface} designed the survey to provide a
fair sampling of the local galaxy population, with as little selection
bias as possible.  Galaxies were chosen without preference for
morphology, environment, inclination, or color, beyond that inherent
in the B-selected and surface-brightness limited parent survey.

An important virtue of the NFGS is its inclusion of the faint galaxy
population.  To counteract the observational bias toward bright
galaxies, \citeauthor{jansen.franx.ea:surface} selected galaxies from
CfA~1 approximately in proportion to the local galaxy luminosity
function \citep[e.g.][]{marzke.huchra.ea:luminosity}.  The resulting
sample spans luminosities from M$_{\rm B}\sim$ $-$23 to $-$15, limited
only by the decreasing CfA~1 survey volume at faint magnitudes.

\section{Data \& Methods}

All distances and magnitudes are computed with H$_{\rm o}$=75, using a
simple Hubble flow model corrected for Virgocentric infall as in
\citet{jansen.franx.ea:surface}.

\subsection{Sample}
The sample analyzed in this paper consists of all 67 NFGS galaxies for
which we have both stellar and ionized gas rotation curve data.  This
group includes 100\% of the E/S0's with detectable ionized gas, i.e.\
17 of 57 or 30\% of all E/S0's.  For later types, the stellar
kinematic database is incomplete: gas and stellar data are available
for 38 of 52 early type Sa-Sbc spirals (73\%), and 12 of 87 later type
spiral and irregular galaxies (14\%).  Figure~\ref{fg:tmb} shows the
sample in the context of the NFGS, demonstrating that the 55 E-Sbc
galaxies we analyze provide a fair sample of the full survey
population of emission line galaxies, whereas the 12 galaxies typed Sc
and later do not.

\subsection{Observations \& Data Reduction}

Long slit spectra were obtained at the F.\ L.\ Whipple Observatory
during several observing runs from 1996 to 1999, primarily using the
FAST spectrograph on the Tillinghast Telescope
\citep{fabricant.cheimets.ea:fast}.  The entire data set and full
details of the data reduction are described elsewhere
\citep{kannappan.fabricant.ea:kinematics}.  In most cases, we have
only major axis spectra.

For the gas kinematics, we observed a $1000$ \AA\ interval centered on
H$_{\alpha}$, with a spectral resolution of $\sigma\sim $ 30 $\rm km
\, s^{-1}$ and a spatial binning of 2.3 arcseconds/pixel, comparable
to the typical seeing of 2''. For the stellar kinematics, we observed
a $2000$ \AA\ interval centered on the MgI triplet at 5175 \AA, again
with 2.3 arcseconds spatial binning, but with spectral resolution
$\sigma\sim $ 60 $\rm km \, s^{-1}$.  We also obtained stellar
kinematic data for a few galaxies with the Blue Channel Spectrograph
on the MMT, using a configuration with 1.2 arcsecond binning, reduced
wavelength coverage, and $\sigma\sim $ 40 $\rm km \, s^{-1}$.  Spectra
of non-rotating G and K giant stars were recorded to serve as velocity
templates for the absorption line data.

All of the data were reduced by standard methods, including bias
and dark subtraction, flat-fielding, wavelength calibration,
heliocentric velocity correction, sky subtraction, spectral
straightening, and cosmic ray removal, using IRAF and IDL.  

We extract high-resolution gas rotation curves (RC's) by
simultaneously fitting H$_{\alpha}$, [NII], and [SII] lines, excluding
data with S/N $<3$.  When possible, we also derive low resolution RC's
from H$_{\beta}$ and [OIII] lines appearing in the stellar absorption
line spectra.  The low resolution RC's act as the primary data for 10
galaxies for which we lack H$_{\alpha}$ observations, and otherwise
serve as confirming data.  In cases of severe H$_{\alpha}$ or
H$_{\beta}$ absorption, we rely on RC's derived from the
[NII], [SII], and [OIII] lines.

To extract stellar rotation curves, we cross correlate the galaxy
absorption line spectra with the stellar template spectra in Fourier
space, using {\bf xcsao} in the rvsao package for IRAF
\citep{kurtz.mink.ea:xcsao}.  We accept fits with R values $> 3.5$ and
error bars $<35$ $\rm km \, s^{-1}$.  We check this procedure by
deriving stellar RC's with a velocity dispersion analysis code kindly
provided by M.\ Franx, described in
\citet{franx.illingworth.ea:major}.  The two methods yield consistent
results within the errors.  We exclude a few points in
the outer parts of one galaxy, A11332\-$+$3536, where the two methods
do not yield overlapping error bars.

\subsection{Identification of Gas-Stellar Counterrotators}

Operationally, we assume that a single rotation curve adequately
describes the kinematics of the stars, and likewise the gas.  In
reality, multiple velocity components may well be present
\citep[e.g.][]{jore:kinematics}, but our spectra have insufficient
signal to noise for us to detect secondary velocity components in an
unbiased way.  Instead we search for bulk counterrotation between the
gas and stars.  Our approach highlights the most extreme
counterrotators, but the number of counterrotators we find should be
considered a lower limit.

We adopt the simplest possible definition of a counterrotator: a
galaxy in which the observed gas and stellar rotation curves show
opposite sign.  Again, this approach yields a lower limit, because our
stellar curves have varying spatial extent, and some galaxies may
contain undetected velocity reversals at large radii, similar to the
reversal we see in NGC~3011 (Figure~\ref{fg:indiv}).  Furthermore, not
all apparent counterrotators actually contain gas and stars in
coplanar counterrotating disks.  An inclined gas disk may also create
an apparent counterrotation signature, and this scenario provides a
plausible interpretation for one of our galaxies (see \S3.1).
However, we note that the processes that produce inclined disks and
those that produce coplanar counterrotating disks are probably
similar, and the former may even evolve into the latter, so our
interpretation does not rest critically on this distinction.

We estimate the confidence level for each counterrotation detection by
attempting to rule out a model in which the stars simply do not
rotate.  Using a standard $\chi^{2}$ minimization algorithm, we fit a
straight line to the stellar rotation curve to determine its slope and
the error on the slope.  In most cases the stellar RC may be
reasonably (if crudely) approximated by a straight line, except in NGC
3011, which shows a mix of corotation and counterrotation (see \S3.1).
In this case we fit just the counterrotating points.  The slope
divided by the error on the slope constitutes the confidence level of
our claim of {\em counter}rotation, as opposed to zero rotation.  For
two galaxies that we label counterrotators, the data differ from a
zero rotation model by only 2--3$\sigma$, so such a model cannot be
completely ruled out.  On the other hand, three S0's and one Im galaxy
rotate so little that we cannot definitely say that they are {\em not}
counterrotators.  We return to this point in \S3.2.
\footnote{Our sample also contains one peculiar case, an Sa galaxy
\citep[NGC~4795, see][]{kannappan.fabricant:frequency} in which the
gas appears to be non-rotating, despite stellar velocities $\sim$150
$\rm km \, s^{-1}$ and despite a tight match between the major axis
position angle and the observed PA's.  There is a faint suggestion
that better data might show counterrotation at radii $\la 1$ kpc, but
the present data do not permit such a claim.  This is the only such
case in our 67 galaxy sample.  NGC~4795 exists in a field of multiple
small companions, and it may be accreting one of them.}

\section{Results \& Discussion}

\subsection{Individual Counterrotators}

\begin{figure}
\epsscale{.84}
\plotone{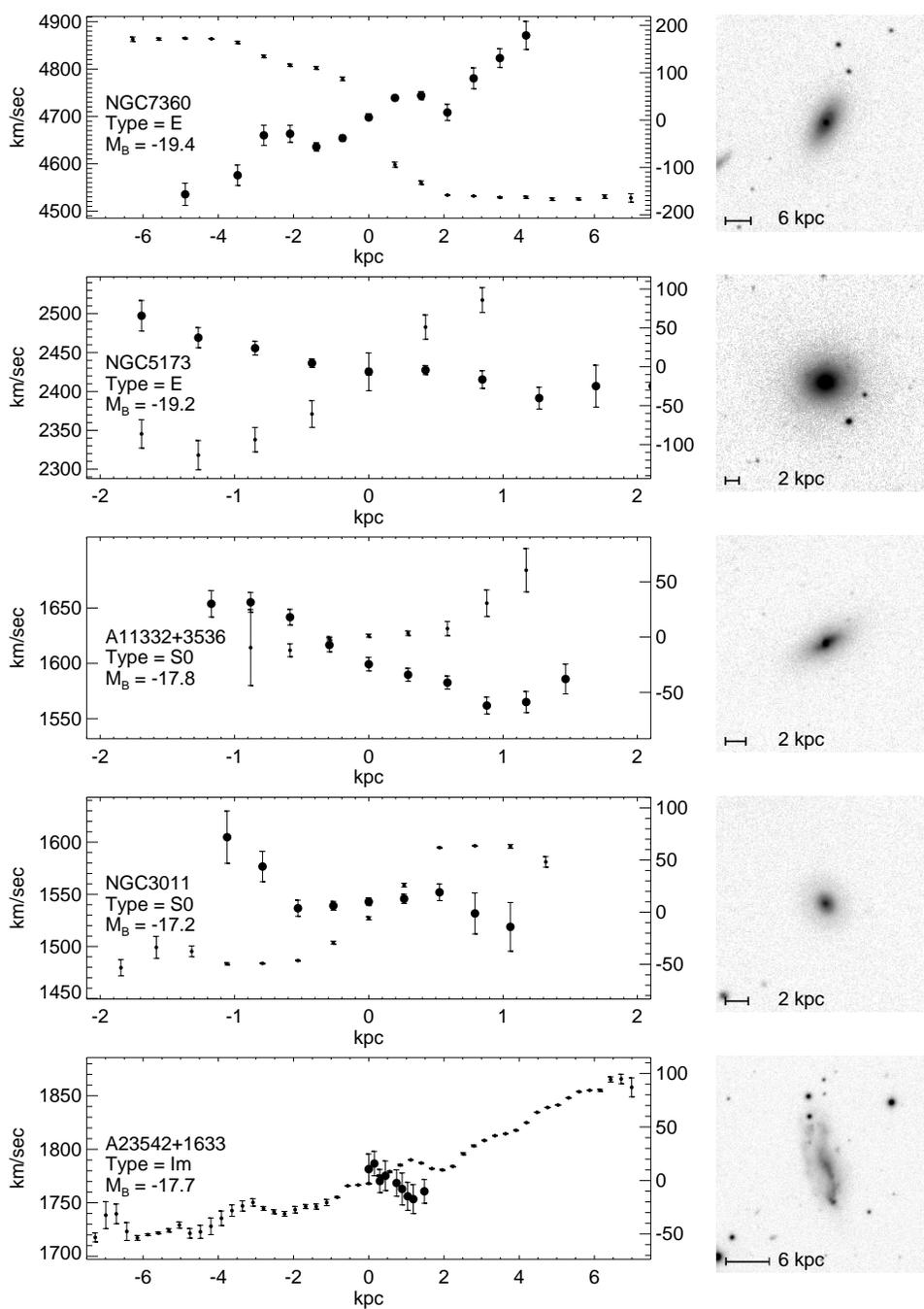}
\caption{Major axis rotation curves and images for the five gas-stellar
counterrotators (see Tables~1--2).  Small dots show ionized gas emission
line data, and large dots show stellar absorption line data.  No
correction has been made for inclination to the line of sight.  Images
shown are U or B-band exposures, courtesy of
\citet{jansen.franx.ea:surface}.}
\label{fg:indiv}
\end{figure}

\begin{figure}
\plotone{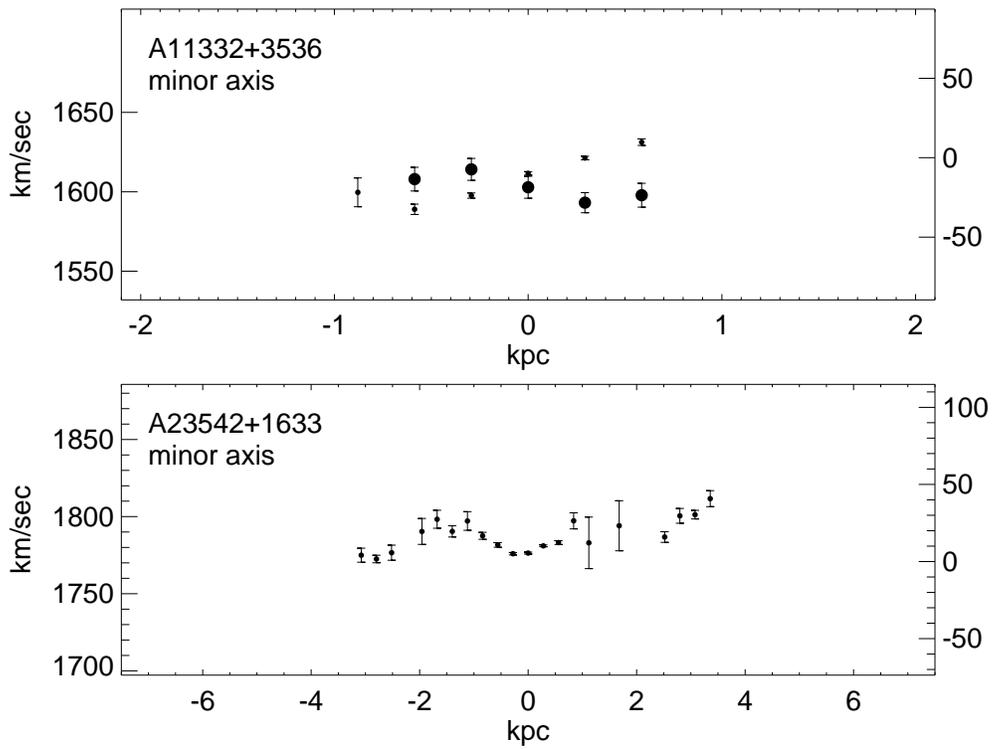}
\caption{Minor axis rotation curves for A11332\-$+$3536 and
A23542\-$+$1633.  Small dots show ionized gas emission line data, and
large dots show stellar absorption line data (none is available for
A23542\-$+$1633).  No correction has been made for inclination to the
line of sight.}
\label{fg:minor}
\end{figure}

In our sample of 67 galaxies, we identify five likely gas-stellar
counterrotators: 2 S0's, 2 E's, and 1 Im.  Figure~\ref{fg:indiv} shows
their rotation curves and images, and Tables 1--4 summarize their
properties.  Figure~\ref{fg:minor} shows the available minor axis data.

We note that the type Im counterrotator found here, A23542\-$+$1633,
may be the first known gas-stellar counterrotator in this morphology
class, although the confidence level of the detection is not decisive
(by fitting the slope as described in \S2.3, we rule out a stellar
non-rotation model at 2.2$\sigma$ confidence, and rule out
equal-amplitude corotation at 3.2$\sigma$ confidence).
\citet{galletta:counterrotation} lists several counterrotators as
``Irr," but these are peculiar rather than Magellanic irregular
galaxies.  The only previous observation of any form of
counterrotation in a Magellanic irregular galaxy appears to be that of
\citet{hunter.wilcots.ea:nature}, who have reported two
counterrotating H\,{\small I} gas components in NGC~4449.  We should
of course consider the possibility that A23542\-$+$1633 might display
misleading velocity reversals due to gas infall along the elongated
structure on its major axis.  However the gas RC does not show large
velocity reversals.  The small velocity shifts near $\sim$$\pm$2 kpc may
reflect infall, especially given the fluctuations seen in the
minor axis RC (Figure~\ref{fg:minor}), but these shifts do not
dominate the major axis kinematics.

In the S0 A11332\-$+$3536, the apparent counterrotation may be caused
by an inclined gas disk.  Consistent with a misaligned disk, the minor
axis gas shows velocity amplitude comparable to the major axis gas,
and while the minor axis gas appears to rotate faster than the stars,
the major axis gas appears to rotate slower than the stars (see
Figures~\ref{fg:indiv} \&~\ref{fg:minor}.  This galaxy also has a
small bar along its minor axis.  \citet{bettoni:on} and
\citet{galletta:counterrotation} point out that radial motions along a
bar may sometimes create a false impression of counterrotation in
one-dimensional data; however, in this case the orientation of the bar
makes confusion due to non-circular motions unlikely.

In NGC~3011, the stellar rotation pattern changes with radius.  Stars
within $\sim$0.5 kpc rotate very slowly (if at all) in the same sense
as the gas, while stars at larger radii counterrotate with 3$\sigma$
confidence.  This combination suggests that the inner stars may
consist of two oppositely rotating stellar populations, one population
having formed more recently from infalling gas.  Further data would be
required to test this hypothesis.

\subsection{Statistics}

\begin{figure}
\plotone{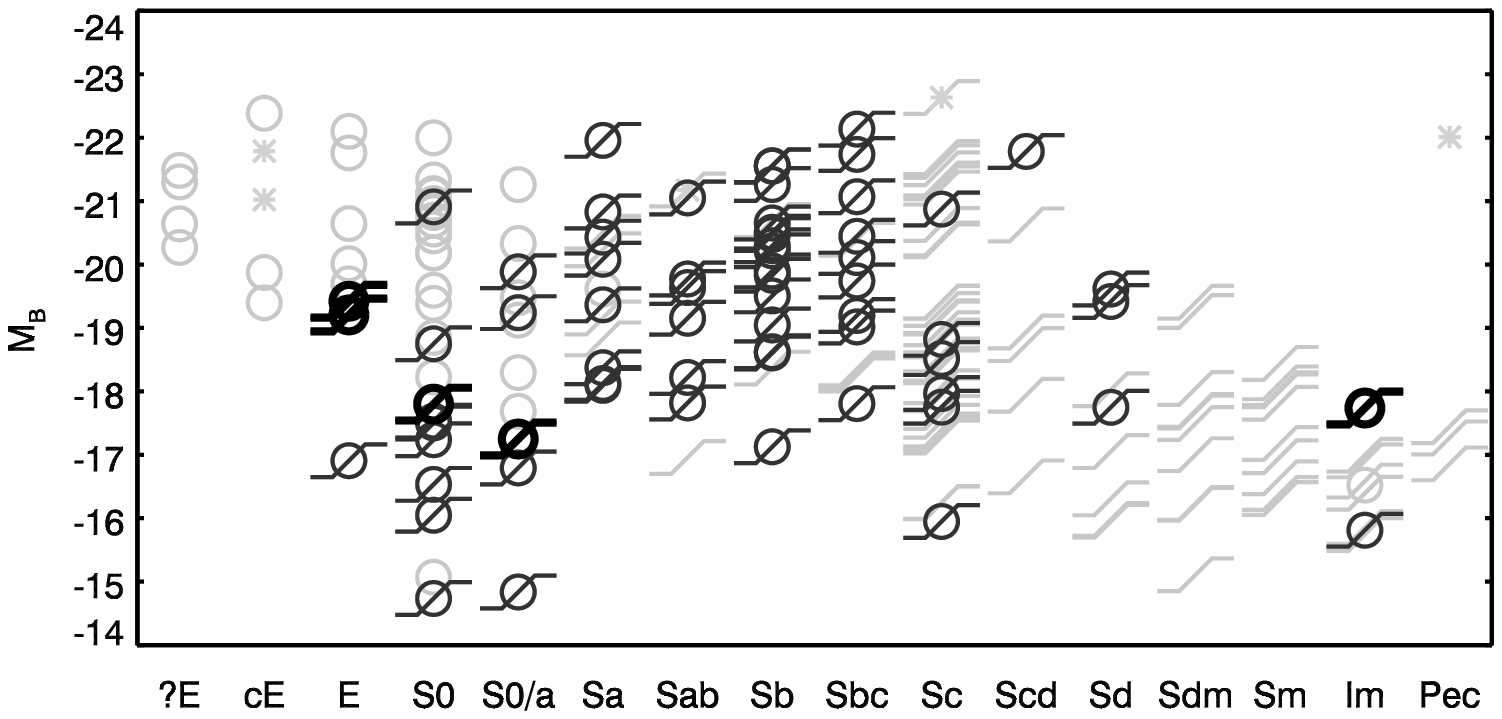}
\caption{An overview of the Nearby Field Galaxy Survey, showing the
demographics of all 196 galaxies sorted by morphology and B-band
luminosity.  A circular symbol indicates that we have stellar
absorption line data for the galaxy, while an S-shaped symbol
indicates that we have extended gas emission line data.  The subsample
of galaxies analyzed for gas-stellar counterrotation consists of all
galaxies for which both symbols appear.  These 67 galaxies are shaded
dark gray, with counterrotators in thick black.  Here we have
separated S0/a's from S0's to reduce crowding in the figure, although
the original morphological classification does not discriminate
reliably between these two classes and all are considered ``S0'' in
the text.  A star indicates that absorption line data were obtained
but are strongly contaminated by an AGN.}
\label{fg:tmb}
\end{figure}

Figure~\ref{fg:tmb} shows the distribution of gas-stellar
counterrotators in morphology and luminosity, within the context of
our 67 galaxy sample as well as the larger NFGS.  The bulk
counterrotation frequency for the 67 galaxy sample is $\sim$7\%.
However, this number includes E/S0's, early type spirals, and later
type galaxies in differing proportions (\S 2.1) and would likely be
lower in a properly weighted sample.

For E/S0's taken alone, gas-stellar counterrotators comprise 4 out of
17 emission line galaxies in the NFGS, which we survey completely.
This result yields a frequency of $24^{+8}_{-6}\%$ (errors are 68\%\
confidence limits from binomial distribution statistics).  If we
exclude the 3 S0's that show almost no rotation (\S2.3), then the
statistics are 4 in 14, or $29^{+10}_{-8}\%$.  In sharp contrast, we
find no clear cases of counterrotation among 38 Sa--Sbc spirals,
although one Sa does show another form of kinematic decoupling (NGC
4795, see note 1).  This rate of non-detection implies that, at 95\%
confidence, no more than $8\%$ of such early type spirals are bulk
counterrotators.  For types Sc and later, we can make no statistical
conclusions due to inadequate sampling.

The strong clustering of the NFGS counterrotators in the early type
morphology region of Figure~\ref{fg:tmb} probably arises from two
factors.  First, if counterrotation originates from galaxy mergers (as
discussed in \S3.3), then early type morphologies are a natural
corollary.  Second, if galaxies of all types were to accrete
retrograde gas with equal probability, then this gas would survive
longer in relatively gas-poor early type galaxies, where collisions
with existing gas and loss of angular momentum are less important.

Figure~\ref{fg:tmb} also illustrates the tendency of our gas-stellar
counterrotators to have low luminosity (sub-L$_{*}$). This result may
simply reflect the fact that low luminosity galaxies, especially early
types, are more likely to be gas rich, while brighter galaxies often
have too little gas for us to detect, and so are not part of our
sample.

Previous studies of gas-stellar counterrotation have focused on
samples restricted by morphology.  \citet{jore:kinematics} and
\citet{haynes.jore.ea:kinematic} analyze the detailed kinematics of a
sample of 23 isolated, unbarred Sa galaxies.  Their sample contains 4
bulk gas-stellar counterrotators, suggesting that we should see
$\sim$1--2 examples among our 8 Sa's.  In fact, we see none, but our
results are consistent within the small number statistics.  Also, we
do find one Sa with kinematically decoupled gas (NGC~4795, see note
1), apparently non-rotating even at radii $\sim$3 kpc.  The sense of
gas rotation in the central $\sim$1 kpc of this galaxy is not well
determined.

In two surveys of S0's with extended gas emission,
\citet{bertola.buson.ea:external} and \citet{kuijken.fisher.ea:search}
independently obtain gas-stellar counterrotation frequencies of
$\sim$20--25\% for samples of 15 and 17 objects respectively.
Combining the two surveys yields a frequency of $\sim$$24^{+6}_{-5}\%$
for a total sample of 29 galaxies, with 3 objects common to both
samples.  Both surveys were drawn from bright galaxy catalogs and
contain objects in the range M$_{\rm B}\sim$ $-$21 to $-$18, with a median
of $\sim$$-$19.

By comparison, the 14 S0's with extended gas emission in the NFGS span
luminosities from M$_{\rm B}=-20.9$ to $-$14.7, with a median of
$\sim$$-$17.  Only $\sim$30\% of these NFGS galaxies overlap the
luminosity range of the two bright galaxy surveys \citep[see
also][]{kannappan.fabricant:frequency}.  Nonetheless, our gas-stellar
counterrotation statistics for these 14 S0's agree with
\citeauthor{bertola.buson.ea:external} and
\citeauthor{kuijken.fisher.ea:search} within the errors:
$14^{+9}_{-6}\%$, or $18^{+12}_{-8}\%$ if we exclude the 3 S0's that
show almost no rotation (\S2.3).

Such agreement suggests the possibility that similar mechanisms form
emission line S0's over a wide range of physical scales --- at least
M$_{\rm B}\sim$ $-$21 to $-$17, the range within which S0 gas-stellar
counterrotators have now been detected.  However, we should point out
that the low luminosity ``S0'' population is heterogeneous, and may
include objects with very different formation histories.  We assigned
the S0 classification purely based on morphology, so it applies to all
NFGS galaxies with the visual appearance of a two-part bulge+disk
structure with minimal spiral structure.  At low luminosities, this
category reveals at least two subclasses: mostly smooth galaxies, and
knotty or otherwise peculiar galaxies, typically with centers bluer
than their outer parts (often labelled blue compact dwarfs).  It is
possible that only the smoother subclass forms a continuum with higher
luminosity S0's in terms of formation history, or alternatively that
the more peculiar objects are simply at an earlier stage of evolution.
Our two S0 counterrotators have most in common with the smoother
subclass, although A11332\-$+$3536 does have a small central bar
and tiny spiral arms.

\subsection{Formation Mechanisms}

The two most plausible formation mechanisms for gas-stellar
counterrotators are late-stage gas accretion and galaxy mergers
\citep[e.g.][and references
therein]{thakar.ryden.ea:ngc,rubin:multi-spin}.  Secular evolution
cannot easily explain large quantities of chemically enriched
counterrotating gas \citep[e.g.][]{caldwell.kirshner.ea:dynamics}.  We
do not separately discuss inclined gas disks here, since they are
likely to be closely related to counterrotating disks.

Late-stage gas accretion mechanisms include acquisition of a large
H\,{\small I} cloud, transfer of gas during a close encounter, and
infall of nearby gas stimulated by a flyby of another galaxy.

Assuming that the counterrotators' H\,{\small I} gas rotates in the same
peculiar sense as their ionized gas, accretion of a single H\,{\small I}
cloud cannot easily explain these galaxies' substantial H\,{\small I} gas
masses ($\sim$$10^{8}-10^{9}$ \MSun, Table~3). If the accreted H\,{\small
I} cloud were similar to the high velocity clouds found near the Milky
Way, then it would have an H\,{\small I} mass of $\sim$$10^{7}$ \MSun, at
least an order of magnitude too small
\citep{blitz.spergel.ea:high-velocity}.  Note that this estimate
assumes that the high velocity clouds are local group objects at $\sim$1
Mpc distances --- if the clouds are closer to the Milky Way, then
their H\,{\small I} masses are even smaller \citep{zwaan.briggs:space}.

The likelihood of either companion gas transfer or infall triggered by
a flyby encounter is harder to evaluate, especially since the
responsible galaxy may have left the neighborhood or may be too faint
to be included in a galaxy catalog.  We can only say that we see no
strong evidence in favor of this scenario.  Based on a search of the
Updated Zwicky Catalog \citep{falco.kurtz.ea:updated} within 600 kpc
and 600 km s$^{-1}$ of each counterrotator, only one counterrotator
has a neighbor within 150 kpc projected on the sky (NGC~5173, see
Table~4).  However, \citet{knapp.raimond:detection} have mapped this
galaxy in detail in H\,{\small I}, and they see no evidence for
H\,{\small I} flow from the companion.  The lack of obviously
interacting companions near our counterrotators is consistent with
expectations based on the analogy between counterrotation and polar
rings.  \citet{brocca.bettoni.ea:visible} compare the local
environments of $\sim$50 apparent polar ring galaxies with those of a
control sample, and they find no statistical difference between the
two groups in the number of close neighbors of comparable luminosity
within $\sim$600 kpc.

\begin{figure}
\epsscale{.7}
\plotone{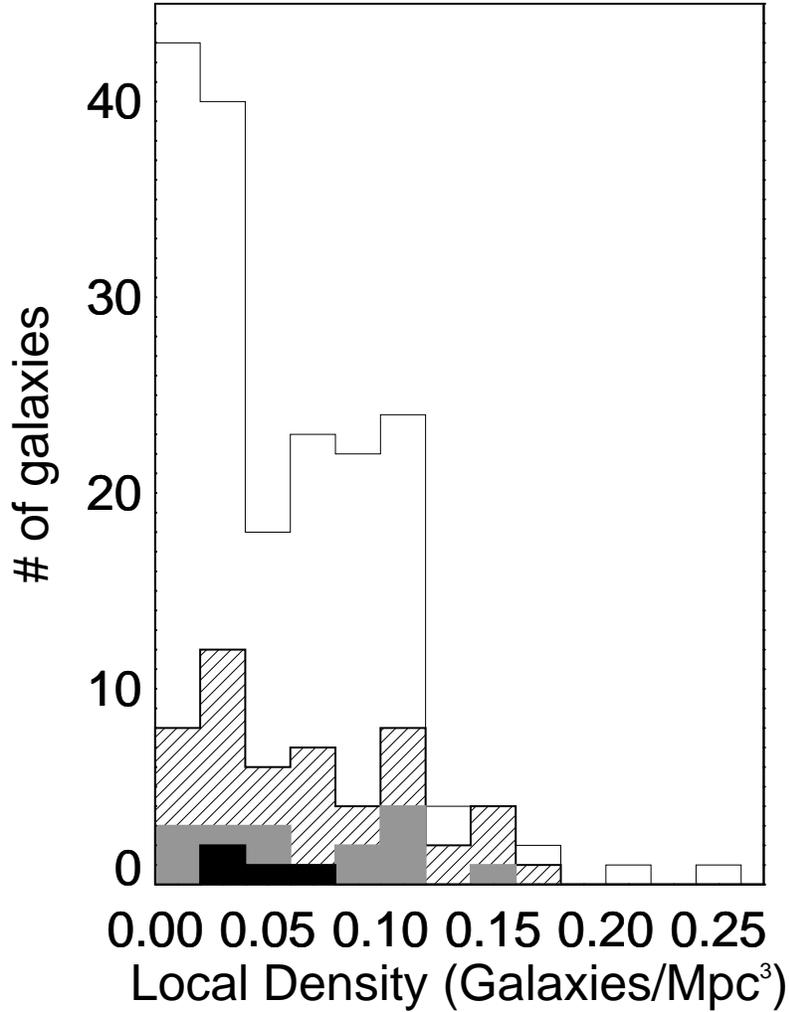}
\caption{A histogram of local galaxy density around each sample
galaxy. Successively smaller histograms represent in order: the entire
NFGS sample, E/S0's in the NFGS sample; E/S0's with extended gas
emission; and the four E/S0 gas-stellar counterrotators.  Densities
were provided by N.\ Grogin (private communication), and were derived
from the CfA~2 survey using the methods described in
\citet{grogin.geller:lyalpha}.  These calculations yield densities
smoothed over $\sim$6.7 Mpc scales (H$_{\rm o}$=75), counting only
galaxies brighter than M$_{\rm B} \sim$ $-$17.
\citeauthor{grogin.geller:lyalpha} compute densities as multiples of
the mean density; we convert these to an absolute scale using their
choice of mean density, 0.03 galaxies/Mpc$^{3}$ for H$_{\rm o}$=75.
Six NFGS galaxies outside the bounds of the CfA~2 survey volume are
excluded from the figure, none of which has extended gas emission.}
\label{fg:dens}
\end{figure}

Of course, the fact that 4 of our 5 counterrotators have E/S0
morphology hints that they probably exist in regions of high local
galaxy density, which density calculations confirm (courtesy
N. Grogin, see Figure~\ref{fg:dens}), although the environments are
only moderately dense.  This observation is consistent with either
flyby or merger formation scenarios.

Mergers provide a simple alternative to late-stage gas accretion
mechanisms.  In this case the companion is gone, so no enhanced
abundance of close neighbors is expected.  The scale of the merger
might range from satellite accretion to a major merger of comparably
sized galaxies (though not necessarily comparably gas rich).  Even
mergers with dwarf galaxies could easily yield the H\,{\small I} gas
masses observed, which are similar to the H\,{\small I} masses of late
type dwarfs in the NFGS \citep[computed from catalog H\,{\small I}
fluxes,][]{bottinelli.gouguenheim.ea:extragalactic,
theureau.bottinelli.ea:kinematics}.\footnote{The relative plausibility
of mergers vs.\ gas accretion would be much better constrained for the
NFGS counterrotators if we could determine the mass and extent of any
counterrotating stellar populations (see comments regarding NGC~3011,
\S3.1).  Our existing data do not permit such an analysis, but we plan
to obtain higher resolution stellar kinematic data to address this
question.}

As discussed in \S3.2, the tendency for bulk counterrotators to have
early type morphologies may mean that both morphology and
counterrotation have a common origin, in which case the merger would
have been substantial.  On the other hand, minor satellite accretion
could also explain the primarily early type morphologies of the
counterrotators, if galaxies of all types accreted small neighbors,
but the retrograde gas did not survive in gas rich later types.  We
expect that retrograde gas will shock with existing prograde gas and
either form stars immediately or lose angular momentum, creating
enhanced infall and central star formation
\citep{lovelace.chou:counterrotating,kuznetsov.lovelace.ea:hydrodynamic}.

For the round elliptical NGC~5173, either scenario is plausible.
\citet{knapp.raimond:detection} propose that this galaxy may have
formed when a gas poor elliptical accreted a gas rich satellite; such
an event could explain the galaxy's apparently small dust to gas mass
ratio \citep{vader.vigroux:star-forming}.  Alternatively, simulations
show that major mergers between gas rich disk galaxies can also
produce counterrotating gas in an elliptical
\citep{hernquist.barnes:origin}.

For the two S0 counterrotators and the disky elliptical NGC~7360,
minor mergers present a likely intermediate option.  Simulations by
\citet{bekki:unequal-mass} and \citet{naab.burkert.ea:on} show that
minor mergers produce disky remnants, and when dissipation and star
formation are included, gas rich minor mergers can produce gas poor
S0's \citep{bekki:unequal-mass}.

\citet{rix.carollo.ea:large} reject the minor merger hypothesis for
low luminosity E/S0's because observations of these galaxies show
significantly greater rotational support than is seen in 3:1 disk
merger simulation remnants.  However, none of the simulations for
which $v/\sigma$ information is available include the physics of
dissipation, infall, and star formation, which are clearly critical
ingredients in the formation of a disk from gas rich progenitors.

In our view, minor or even gas-rich major mergers provide a very
plausible formation mechanism for disky low luminosity E/S0's, if one
considers that the likely progenitors bear little resemblance to the
model galaxies currently used in merger simulations.
\citet{mihos.hernquist:ultraluminous} have shown that bulgeless disk
galaxy mergers evolve very differently from bulge + disk galaxy
mergers, but no one has yet modelled the complexities of a dwarf-dwarf
merger.  For example, the progenitors might be gas rich dwarfs with
extended filamentary H\,{\small I} envelopes that continue to fall
into the remnant late in the merger process.  The gas dynamics of
small gravitational potential wells may also be important: many small
early types in the NFGS display broad emission line wings, possibly
related to mass outflows.

Given the steeply rising numbers of galaxies at the faint end of the
galaxy luminosity function \citep{marzke.huchra.ea:luminosity}, gas
rich dwarf-dwarf mergers are inevitable, and some of the remnants of
such mergers may well look like S0's, or spirals embedded in S0
envelopes.  The existence of very low luminosity S0's, and
counterrotators as faint as M$_{\rm B}\sim$ $-$17, contrasts with the
sharp decrease in early type spirals fainter than M$_{\rm B}\sim$ $-$18
and the corresponding increase in the number of late type dwarfs
\citep[Figure~\ref{fg:tmb}, see
also][]{sandage.binggeli.ea:studies,schombert.pildis.ea:dwarf,marzke.huchra.ea:luminosity}.
We note that the faint population of the NFGS contains a number of
possible ``proto-S0's'' --- galaxies with very blue centers and outer
envelopes reminiscent of early types \citep{jansen.franx.ea:surface}.
These galaxies are variously typed late or early depending on the
surface brightness of the envelope and the degree of inner structure.
In the nomenclature of the dwarf galaxy literature, some would be
known as blue compact dwarfs, a class of galaxies showing many of the
expected characteristics of merger remnants
\citep{doublier.comte.ea:multi-spectral}.
\footnote{ None of this precludes that some S0's may form through
tidal stripping, but it would be rather difficult to form a
counterrotator that way, unless the formation of the galaxy and the
acquisition of counterrotating gas were entirely separate events. }

Of course, some dwarf merger remnants probably just turn into bigger,
brighter late type dwarfs.  For example, although the Im
counterrotator A23542\-$+$1633 appears to be actively evolving, with
ongoing infall (\S3.1) and moderately strong star formation (Table~3),
it does not much resemble a proto-S0, but appears more like a
proto-Sd.  

If the counterrotators are merger remnants, one might expect them to
show signatures of enhanced past or present star formation.  On the
other hand, the E/S0 counterrotators' smooth morphologies (with small
perturbations, see Table~1) suggest that for these four galaxies, the
merger probably took place at least $\sim$1 Gyr ago
\citep{schweizer:observational}.  In keeping with this view, these
galaxies show moderate H$_{\delta}$ absorption equivalent widths
($\sim$1--3 \AA, R.\ Jansen, private communication), consistent with
starburst ages greater than $\sim$1.5 Gyr for solar metallicity
\citep{worthey.ottaviani:h}.  By contrast, the one irregular
counterrotator shows stronger H$_{\delta}$ absorption ($\sim$5--6 \AA,
subject to some uncertainty in the emission correction), suggesting a
more recent/ongoing starburst.  We note also that for the two faint
S0's, the nominal gas consumption timescale is relatively short ($\sim$2 
Gyr uncorrected for recycling, see Table~3), possibly implying
rapid evolution.  In fact, both of these galaxies are Markarian
galaxies and have starburst nuclei \citep{balzano:star-burst}.

\section{Conclusions}

We have searched for bulk gas-stellar counterrotation in 67 galaxies
spanning a broad range of morphologies and luminosities within the
$\sim$200 galaxy NFGS sample.  This sample permits statistical
conclusions for types E--Sbc, and includes a few later types as well.
However, our detections represent a lower limit to the true rate of
counterrotation, because the data do not permit separation of multiple
kinematic components and do not rule out counterrotation beyond the
radial extent of the observations.

We detect 5 gas-stellar counterrotators, generally of early type and
low luminosity.  These galaxies include 2 E's, 2 S0's, no spirals, and
1 Magellanic irregular.  The Im galaxy counterrotates with 2.2$\sigma$
confidence, and if confirmed represents the first known example of
gas-stellar counterrotation in a Magellanic irregular.  One of the S0
counterrotators probably contains an inclined gas disk rather than
coplanar counterrotation; we assume that these phenomena are closely
related in our interpretation.

Statistically, we conclude that $24^{+8}_{-6}\%$ of E/S0's with
extended emission are bulk gas-stellar counterrotators, or
$29^{+10}_{-8}\%$ if we exclude 3 S0's that display very little
rotation.  In contrast, our non-detection of spiral counterrotators
implies that no more than 8\% of Sa--Sbc spirals are bulk
counterrotators, at 95\% confidence.  This morphological dependence of
counterrotation frequency may arise from two effects.  First, if
galaxy interactions and mergers are responsible for creating the
counterrotators, then the same mechanisms will tend to produce early
type morphologies.  Second, even small-scale retrograde gas accretion
events that do not strongly reshape morphology will be easier to
detect in E/S0's, because bright spiral galaxies will typically have
sufficient prograde gas to dynamically neutralize the infall.

Sa galaxies may represent a transitional case.  Although we detect no
Sa counterrotators, our sample is small (only 8 galaxies) and includes
one galaxy in which the sign of the central gas rotation is uncertain.
The same galaxy shows clear gas-stellar decoupling at larger radii,
where its gas displays zero apparent rotation, despite large stellar
velocities.  \citet{jore:kinematics} finds that $\sim$15--20\% of
bright, unbarred Sa's are bulk gas-stellar counterrotators.

For S0's with extended gas emission, the frequency of gas-stellar
counterrotation we derive agrees with the results of
\citet{bertola.buson.ea:external} and
\citet{kuijken.fisher.ea:search}, although the median luminosity of
our sample is $\sim$2 mag fainter (M$_{\rm B}\sim$ $-$17).  The agreement
suggests that similar mechanisms form this category of S0's over a
wide range of physical scales, at least $-17>\rm M_{\rm B} >-21$.

As noted by \citet{bertola.buson.ea:external}, every known
counterrotator implies at least one corotator that formed by a similar
process, and probably more due to effects that tend to erase the
retrograde kinematic signature.  Therefore the $\sim$25\% bulk
counterrotation rate for emission line E/S0's implies that at least
$\sim$50\% of E/S0's with extended gas emission have experienced the
evolutionary processes that produce gas-stellar counterrotation.

In examining the range of possible processes, we conclude that galaxy
mergers (including satellite accretion) provide the most plausible
explanation for the counterrotators, especially given these galaxies'
significant H\,{\small I} masses and lack of obvious companions.
However flyby or faint companion interactions remain a possibility.
We note that the possible products of gas rich dwarf-dwarf mergers
remain largely unexplored in detailed simulations, despite clear
evidence for such mergers in the faint galaxy population, and we
suggest that our S0 and Im dwarf counterrotators may be products of
such mergers.

\acknowledgments 

Rolf Jansen generously provided his data and assisted us in using it.
Norm Grogin kindly calculated local galaxy densities for us.  Nelson
Caldwell, Marijn Franx, and Lars Hernquist made helpful suggestions.
Finally, Betsy Barton, Barbara Carter, Emilio Falco, Martha Haynes,
John Huchra, Bob Kirshner, Douglas Mar, Hans-Walter Rix, and Aaron
Romanowsky all provided information or resources for which we are
extremely grateful.  S.~J.~K. acknowledges support from a NASA GSRP
Fellowship.

\newpage

\newpage

\newpage

\newpage

\newpage
\clearpage

\renewcommand{\arraystretch}{.6}

\begin{deluxetable}{llccccccl}
\tablenum{1} 
\tabletypesize{\footnotesize}
\tablewidth{0pt}
\tablecaption{Basic Properties of Counterrotators
\label{tab1}}
\tablehead{\colhead{Galaxy} & 
\colhead{UGC\#} &
\colhead{Confidence\tablenotemark{a}} &
\colhead{Distance\tablenotemark{b}} &
\colhead{M$_{\rm B}$\tablenotemark{b}} &
\colhead{r$_{e}$\tablenotemark{b}} &
\colhead{Type\tablenotemark{c}} & 
\colhead{Morphology Notes} \\
& 
&
\colhead{($\sigma$)} & 
\colhead{(Mpc)} & 
\colhead{(mag)} & 
\colhead{(kpc)} & 
& 
} 
\startdata 
    NGC 3011 & 5259 & \phn3.0 & 24.0 & -17.2 & 0.8 & S0 & completely smooth      \\
 A11332\-$+$3536 & 6570 & 11.6 & 26.7 & -17.8 & 1.2 & S0 & faint inner bar+arms   \\
    NGC 5173 & 8468 &  \phn5.9 & 38.5 & -19.2 & 1.8 & E  & outer knottiness, arms\tablenotemark{d}  \\
    NGC 7360 & 12167 & 17.5 & 63.4 & -19.4 & 3.9 & E  & smooth, very elongated \\
 A23542\-$+$1633 & 12856 &  \phn2.2 & 25.4 & -17.7 & 4.0 & Im & knotty elongated core  \\
\enddata

\tablenotetext{a}{Confidence level of claim that stars {\em
counter}rotate, as opposed to having zero rotation. For NGC~3011, only
the outer four points are considered.  See \S2.3.}
\tablenotetext{b}{Distances, magnitudes, and effective radii from
\citet{jansen.franx.ea:surface}, converted to H$_{\rm o}$=75.}
\tablenotetext{c}{Morphological types as used by
\citet{jansen.franx.ea:surface}, except that we refer to all
lenticular galaxies from L- to S0/a as ``S0.''}
\tablenotetext{d}{\citet{vader.vigroux:star-forming} observe possible
spiral arms in their continuum-subtracted B-band image of this
galaxy.}
\end{deluxetable}

\begin{deluxetable}{llll}
\tablenum{2}
\tabletypesize{\footnotesize}
\tablewidth{0pt}
\tablecaption{Position Angles Observed
\label{tab2}}
\tablehead{\colhead{Galaxy} &
\colhead{Major Axis PA~\tablenotemark{a}} & 
\colhead{Stellar RC PA} &
\colhead{Gas RC PA} } 
\startdata 
    NGC 3011  & 52 & 50 & 52 \\
    A11332\-$+$3536  & 123 & 122, 32 (minor axis) & same observations~\tablenotemark{b} \\
    NGC 5173  & 100  & 100 & same observation~\tablenotemark{b} \\
    NGC 7360  & 153 & 153 & 153 \\
    A23542\-$+$1633  & 12 & 20~\tablenotemark{c} & 12, 102 (minor axis) \\
\enddata
\tablenotetext{a}{Major axis PA's are as compiled by
\citet{jansen.franx.ea:surface} from the Uppsala General Catalog of
Galaxies \citep[UGC,][]{nilson:uppsala}; except for NGC~5173, for
which the UGC lists indefinite PA.  \citet{vader.vigroux:star-forming}
measure a PA of $100\pm5$ from their high-quality B band image of this
galaxy.}
\tablenotetext{b}{For A11332\-$+$3536 and NGC~5173, we do not have
data in the 6000--7000 \AA\ range, so our primary gas RC's are derived
from [OIII] and H$_{\beta}$ emission lines in the stellar absorption
line spectra.}  
\tablenotetext{c}{This observation was obtained at the Multiple Mirror
Telescope Observatory, a facility operated jointly by the University
of Arizona and the Smithsonian Institution.}

\end{deluxetable}

\begin{deluxetable}{lcccccc}
\tablenum{3} 
\tabletypesize{\footnotesize}
\tablewidth{0pt}
\tablecaption{Gas and Star Formation Properties of Counterrotators
\label{tab3}}
\tablehead{\colhead{Galaxy} &
\colhead{$M_{\rm HI}$~\tablenotemark{a}} & 
\colhead{${\rm M_{\rm HI}}/{\rm L_{\rm B}}$} &
\colhead{SFR\tablenotemark{b}} &
\colhead{T\tablenotemark{c}} &
\colhead{EW(H$_{\alpha}$)\tablenotemark{d}} &
\colhead{$B-R$\tablenotemark{e}} \\
& 
\colhead{(\MSun)} & 
\colhead{(\MSun$/$\LSun)} & 
\colhead{(\MSun/yr)} & 
\colhead{(Gyr)} & 
\colhead{(\AA)} & 
\colhead{(mag)} 
} 
\startdata 
    NGC 3011 & 8.6E+07 & 0.07 & 0.055 & 1.6  & -15.81 & 1.20 \\
 A11332\-$+$3536 & 2.4E+08 & 0.12 & 0.096 & 2.5  & -15.42 & 1.17 \\
    NGC 5173 & 2.1E+09 & 0.28 & 0.078 & 27 & \phn-3.13 & 1.29 \\
    NGC 7360 & 3.6E+09 & 0.39 & 0.073 & 49 & \phn-2.37 & 1.35 \\
 A23542\-$+$1633 & 2.3E+09 & 1.21 & 0.220 & 11 & -62.33 & 0.69 \\
\enddata
\tablenotetext{a}{Computed from H\,{\small I} fluxes
\citep{bottinelli.gouguenheim.ea:extragalactic,theureau.bottinelli.ea:kinematics}.}
\tablenotetext{b}{Computed from H$\alpha$ fluxes 
(R.\ Jansen, private communication) using the calibration of \citet{kennicutt:star}.}
\tablenotetext{c}{Gas consumption timescale uncorrected for recycling.}
\tablenotetext{d}{\citet{jansen.fabricant.ea:spectrophotometry}.}
\tablenotetext{e}{\citet{jansen.franx.ea:surface}.}
\end{deluxetable}

\begin{deluxetable}{lll}
\tablenum{4}
\tabletypesize{\footnotesize}
\tablewidth{0pt}
\tablecaption{Neighbors of Counterrotators~\tablenotemark{a}
\label{tab4}}
\tablehead{\colhead{Galaxy} &
\colhead{Nearest Neighbor} & 
\colhead{Nearest Brighter Neighbor} } 
\startdata 
    NGC 3011  & 270 kpc, 0.6 mag fainter & 440 kpc, 2 mag brighter\\
 A11332\-$+$3536  & 175 kpc, 0.5 mag fainter & 240 kpc, 0.4 mag brighter\\
    NGC 5173  & 50 kpc, 1.2 mag fainter  & 180 kpc, 0.3 mag brighter\\
    NGC 7360  & none in UZC\tablenotemark{b} & \\
 A23542\-$+$1633  & 480 kpc, 0.1 mag brighter\tablenotemark{c} & same as nearest neighbor\\
\enddata
\tablenotetext{a}{Based on a search of the Updated Zwicky Catalog
\citep{falco.kurtz.ea:updated} 
within 600 kpc and 600 km s$^{-1}$; all
neighbors listed have $\Delta$v$<$250 km s$^{-1}$.}
\tablenotetext{b}{Small projected neighbor appears to be a background galaxy.}
\tablenotetext{c}{A bright knot within this galaxy has at times been interpreted as a neighbor, but is in fact part of the main galaxy.}
\end{deluxetable}

\end{document}